\documentclass[sigconf]{nimeart}
\usepackage{balance}
\AtBeginDocument{%
  }

\setcopyright{cc}
\copyrightyear{2026}
\acmYear{2026}
\acmDOI{}

\acmConference[NIME '26]{International Conference on New Interfaces for Musical Expression}{June 23--26,
  2026}{London, UK}
\acmISBN{}
\begin{document}

\title{PHOTON: Non-Invasive Optical Tracking of Key-Lever Motion in Historical Keyboard Instruments}
\author{Noah Jaffe}
\orcid{0009-0007-7659-967X}
\email{n.jaffe@uva.nl}
\affiliation{%
  \institution{Institute for Logic, Language, and Computation}
  \institution{University of Amsterdam}
  \city{Amsterdam}
  \country{The Netherlands}
}

\author{John Ashley Burgoyne}
\email{j.a.burgoyne@uva.nl}
\affiliation{%
  \institution{Institute for Logic, Language, and Computation}
  \institution{University of Amsterdam}
  \city{Amsterdam}
  \country{The Netherlands}
}

\renewcommand{\shortauthors}{Jaffe and Burgoyne}

\begin{abstract}
This paper introduces PHOTON (PHysical Optical Tracking of Notes), a non-invasive optical sensing system for measuring key-lever motion in historical keyboard instruments.
PHOTON tracks the vertical displacement of the key lever itself, capturing motion shaped by both performer input and the instrument’s mechanically imposed, time-varying load.
Reflective optical sensors mounted beneath the distal end of each lever provide continuous displacement, timing, and articulation data without interfering with the action.

Unlike existing optical systems designed for modern pianos, PHOTON accommodates the diverse geometries, limited clearances, and non-standard layouts of harpsichords, clavichords, and early fortepianos.
Its modular, low-profile architecture enables high-resolution, low-latency sensing across multiple manuals and variable key counts.
Beyond performance capture, PHOTON provides real-time MIDI output and supports empirical study of expressive gesture, human–instrument interaction, and the construction of instrument-specific MIDI corpora using real historical mechanisms.

The complete system is released as open-source hardware and software—from schematics and PCB layouts developed in KiCad to firmware written in CircuitPython—lowering the barrier to adoption, replication, and extension.
\end{abstract}


\begin{teaserfigure}
  \includegraphics[width=\textwidth]{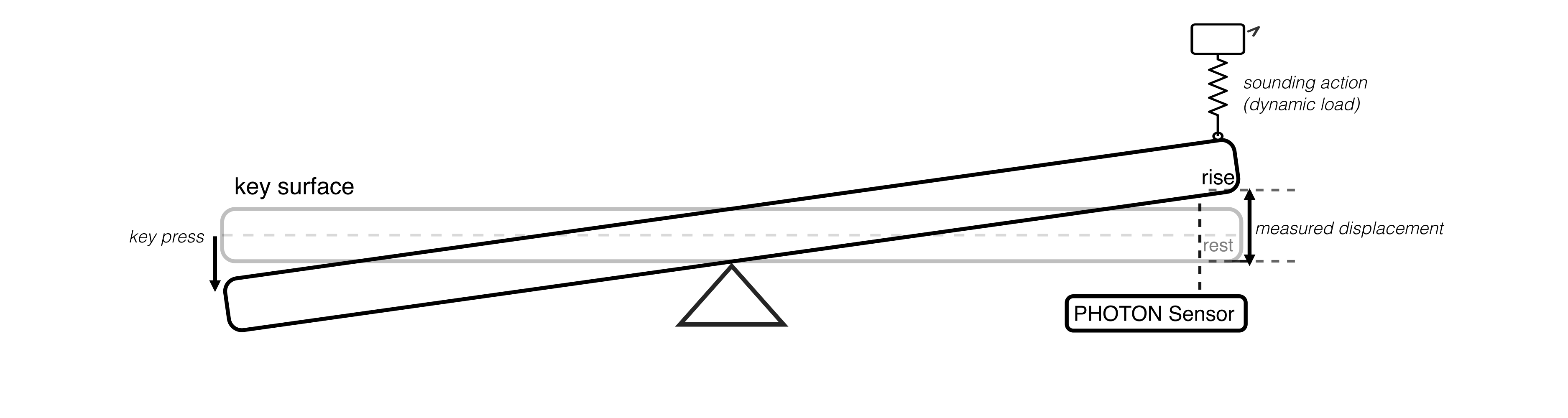}
  \caption{Measurement of key-lever motion under dynamic sounding load.}
    \Description{Diagram of a keyboard key lever showing how the PHOTON system measures motion. A reflective optical sensor is mounted beneath the distal end of the key lever and measures vertical displacement as the key is pressed and released. The lever pivots around a balance point, and the influence of the instrument’s sounding action is represented as an abstract dynamic mechanical load acting on the lever.}
  \label{fig:keylever}
\end{teaserfigure}

\maketitle

\section{Introduction}
This work introduces PHOTON (PHysical Optical Tracking of Notes), a modular, non-invasive reflective optical key-tracking system designed specifically for historical keyboard instruments. Although developed for historical keyboards, the architecture is applicable to a wide range of keyboard mechanisms.
PHOTON measures the motion of the key lever itself, rather than attempting to sense sound-producing components directly. The displacement, velocity, and timing of the lever are jointly determined by the performer’s physical gesture and by the resistance, inertia, and return forces imposed by the instrument’s action, which differ fundamentally across keyboard instruments.
In the piano, the key lever interacts with a complex escapement and hammer mechanism; in the harpsichord, it lifts a jack whose plectrum presses against the string before releasing it; and in the clavichord, it drives a tangent that remains in contact with the string.
As a result, the motion measured at the sensing point emerges from the continuous interaction between the performer’s gesture and the instrument’s mechanical action.

\begin{figure}[ht]
    \centering
    \includegraphics[width=1.0\linewidth]{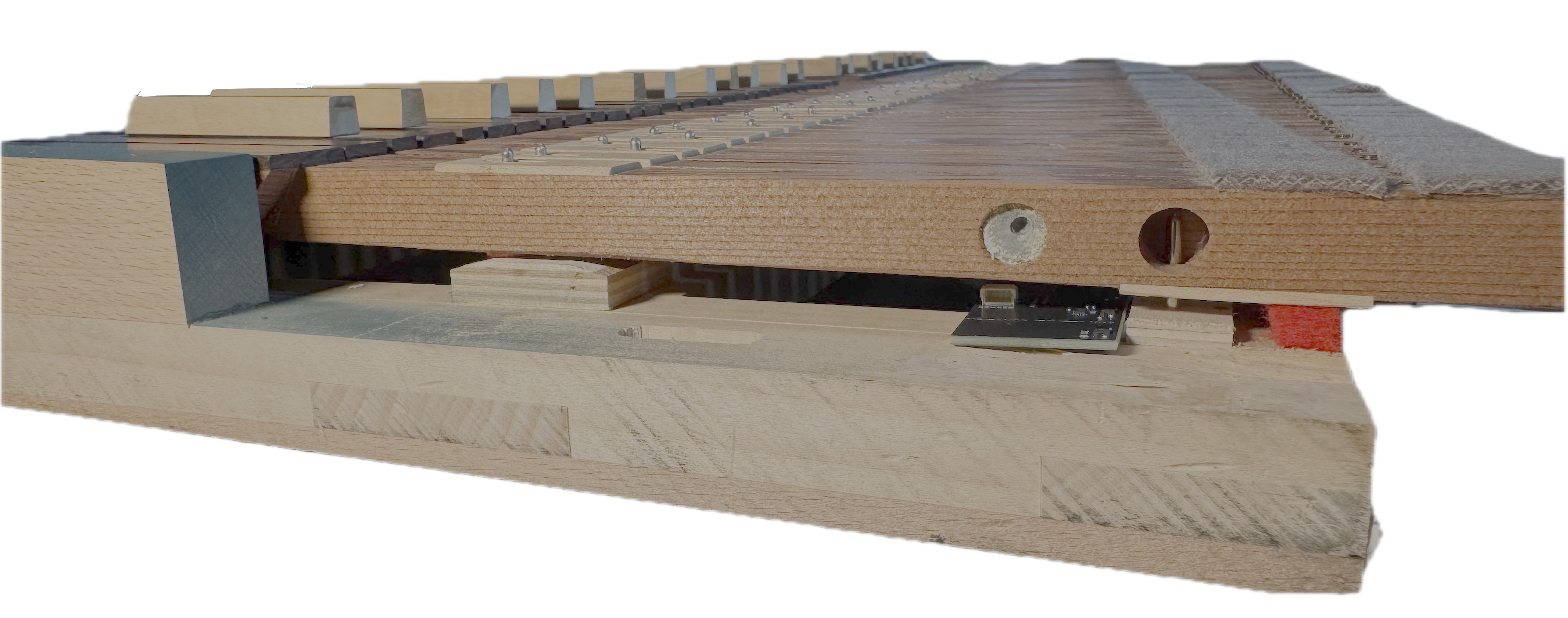}
    \caption{Harpsichord manual with PHOTON module.}
    \Description{Photograph of a detached harpsichord manual viewed from the right, showing the full row of wooden keys with a PHOTON sensing module mounted under the rear of the key levers.}
    \label{fig:photon_installed}
\end{figure}

By positioning a reflective optical sensor beneath the distal end of the key lever, PHOTON captures this mechanically mediated motion without interfering with the action, measuring vertical motion continuously throughout each keystroke.
An example installation geometry is shown in Figs.~\ref{fig:keylever} and ~\ref{fig:photon_installed}.  
Because the sensing location lies within the mechanically coupled chain between the performer’s finger and the sound-producing elements, the resulting signal encodes both gestural intent and the instrument’s mechanically mediated response. 

\subsection{Related Work and Problem Statement}
Recent related work has sensed motion at different locations and with different system architectures.
Hamilton et al.\ detect harpsichord jack motion using gradient markers placed on the sides of the jacks, thereby capturing the moment at which string excitation occurs~\cite{hamilton2025harpsichord}.
In contrast, PHOTON measures key-lever motion from beneath the manual using a lower-profile and less invasive arrangement that is more readily extended to double-manual instruments; it also avoids the need for optical baffles between adjacent sensors and is compatible with substantially tighter installation spaces.
More closely related to the present work, Schmidt et al.\ place sensors beneath the distal ends of the keys to capture key position~\cite{schmidt2025sparksichord}.
PHOTON adopts the same general sensing location, but with a different hardware architecture intended for tighter vertical clearances and simpler interconnection, as discussed in Section~\ref{sec:system_integration}.
PHOTON also differs in acquisition strategy: sensors are enabled selectively rather than driven continuously, which supports scaling to larger sensor counts. 
For example, a five-octave two-manual harpsichord requires 122 sensors.
Taken together, these prior systems demonstrate the feasibility of optical sensing on keyboard instruments, but do not directly address the spatial, mechanical, and deployment constraints posed by historical keyboards, which PHOTON is designed to accommodate.

Most other prior work on non-contact key sensing has focused on the modern piano.
Commercial non-contact sensing systems for modern pianos—most notably QRS Music Technologies’ PNOScan and the now-discontinued Moog PianoBar—have existed for decades, supporting MIDI performance capture, automated playback, and integration with digital audio workstations (DAWs) \cite{Mowat2005MoogPianoBar,QRS}.
Research systems have likewise demonstrated the feasibility and musical value of continuous key-motion sensing on acoustic pianos. 
In particular, \citeauthor{McPherson2013ContinuousPianoGesture} introduced a portable optical system for measuring continuous key position on any piano, enabling detailed study and mapping of expressive touch beyond conventional MIDI representations \cite{McPhersonKim2011CHI, McPherson2013ContinuousPianoGesture}. 
More recent work, such as the system introduced by \citeauthor{Oku2022}, further demonstrates that optical key tracking can resolve fine-grained expressive detail in piano performance \cite{Oku2022, kuromiya2025motor}.
However, both commercial and research systems of this kind are designed around the standardised geometry of the modern piano, whose 88-note compass, dimensions, and layout have remained consistent for more than a century.

Historical keyboard instruments, by contrast, vary in compass and key dimensions, and may include alternative layouts such as short octaves or split keys.
As a result, modern piano-oriented sensing systems cannot be directly installed in a harpsichord, clavichord, or fortepiano. 
In addition, existing solutions are typically proprietary or bespoke research platforms, limiting their adaptability and availability for wider use.
Beyond geometric incompatibility, historical keyboards impose a distinct set of mechanical and practical constraints.
Clearances within the action and keybed are often minimal, and any sensing system must be installed non-invasively, without permanent modification to the instrument.
Consequently, only a very small number of MIDI-enabled harpsichords have existed in research contexts, such as the instrument used by \citeauthor{Gingras2013IndividualityHarpsichord} to study voice emphasis, timing, articulation, and velocity in professional performance \cite{Gingras2009VoiceEmphasisHarpsichord, Gingras2013IndividualityHarpsichord}. Recent computational work has begun to address early keyboard practice directly. 
\citeauthor{Stefunko2025ACoRD} introduced ACoRD,\footnote{\url{https://lindat.mff.cuni.cz/repository/items/26323f0c-718a-4f4f-bf0b-396accd46a30}} the first symbolic performance dataset of human-performed basso continuo realizations, comprising 175 MIDI recordings by professional and student harpsichordists \cite{Stefunko2025ACoRD}. 
Subsequent work has proposed feature representations and alignment-based analyses on this dataset to study individuality and stylistic variation in basso continuo performance \cite{Stefunko2026Griff}.
Together, this work demonstrates the analytical value of symbolic continuo performance corpora derived from generic MIDI controllers.
By design, micro-timing, dynamics, and mechanically mediated aspects of touch and articulation are not represented, leaving expressive gesture on historical keyboards largely unexplored.

\subsection{Motivation}
To enable empirical study of touch, articulation, and expressive gesture on historical keyboards, this work introduces a sensing architecture that is inexpensive to manufacture, non-invasive to install, adaptable to diverse keyboard layouts, and discreet in operation. 
The system is modular, low-profile, and entirely non-contact, providing continuous access to key-lever behaviour without altering the instrument.
It produces MIDI event streams suitable for digital audio workstations, virtual instruments, and large-scale performance analysis. 
Empirical music performance research has historically focused on the modern piano, in part because it was among the first instruments to benefit from widely available MIDI-enabled infrastructures, leaving historical keyboard performance comparatively underrepresented in empirical research \cite{Gabrielsson2003PerformanceMillennium}. 
Beyond its immediate role as a sensing device, PHOTON addresses the lack of empirical access to expressive gesture on historical keyboard instruments.

In practical terms, the project also functions as a reusable reference design for building historical keyboard–computer interfaces. While basic electronics assembly and software skills are still required, the availability of validated architecture, schematics, PCB layouts, bills of materials, and firmware substantially reduces the effort required to retrofit a historical keyboard.

\section{Applications}
PHOTON enables a range of analytical, artistic, and interactive applications for historical keyboard instruments.
In most cases, it functions as part of a larger system, providing motion or performance data that are interpreted or rendered by external software, virtual instruments, or mapping layers.
The following subsections describe representative configurations of such systems.

\subsection{Motion Capture}
PHOTON provides continuous, high-resolution measurements of key travel over time, enabling direct visualisation and analysis of key-motion trajectories. These time–displacement traces reveal fine-grained aspects of touch and articulation—such as attack profiles, aftertouch behaviour, release dynamics, and mechanical settling—that are not accessible through audio recordings or discrete MIDI note events alone. 
Example motion traces captured with PHOTON are presented in Section~\ref{sec:case_studies} as part of a case study examining harpsichord key–action behaviour.

The system yields approximately 100 reliably distinguishable position levels over the relevant keystroke (approximately 1~cm of travel).
These levels are derived from per-sensor ADC readings calibrated between two mechanically measured reference positions, and provide a high-resolution, monotonic estimate of key-lever displacement.
Because the reflective sensor response is not intrinsically linear with distance, the intermediate values should be interpreted as nonlinearly interpolated coordinates rather than as absolute geometric measurements.
In addition to full-array scanning, PHOTON can be configured to continuously monitor a selected subset of sensors, allowing substantially higher temporal resolution when required.
In this mode, selected sensors can be sampled at rates in excess of 250\,Hz.

Together, these motion traces and measurement capabilities support empirical studies of keyboard technique, performer\allowbreak--\allowbreak instrument interaction, and historically informed performance practice.
For performers and teachers, they offer a concrete means of examining and comparing technical gestures. For musicologists and organologists, they enable quantitative comparison of how different historical actions shape the physical realisation of musical intent.

In the current implementation, PHOTON presents itself as a composite USB device over a single connection.
The USB architecture supports both standard MIDI note-event communication and higher-resolution position streaming via serial, but the present firmware exposes these in separately selectable software modes rather than as simultaneous data streams.
This preserves compatibility with existing MIDI workflows while providing access to continuous motion data for analysis and custom applications.

\subsection{Live Sound Reinforcement}
Beyond analytical applications, PHOTON also supports real-time performance contexts 
involving digital sound reinforcement. In such configurations, a harpsichord or 
fortepiano can drive a virtual instrument in parallel with its acoustic action, with 
the resulting signal sent to a loudspeaker or transducer system. This allows the 
instrument's sound to be reinforced and projected beyond its natural volume while 
preserving its characteristic timbre. 
Because the virtual instrument is driven independently of the acoustic signal, this 
approach also offers practical benefits that microphone-based reinforcement cannot 
provide — for example, the tuning stability of a digital instrument in live performance contexts.

\subsection{Hyperinstruments}
Machover and Chung introduced the term \emph{hyperinstrument} to describe instruments augmented with sensing and computation to extend their expressive and interactive capabilities beyond those of the acoustic instrument alone \cite{MachoverChung1989Hyperinstruments}.
Recent work on the \emph{Global Hyperorgan} shows how pipe organs equipped with networked control can function as hyperinstruments, enabling telematic performance and the real-time coupling of physically separate organs into a single extended instrument \cite{Harlow2021GlobalHyperorgan}. 

PHOTON enables hyperinstrument configurations by letting historical keyboards control other digitally enabled instruments via MIDI.
As a concrete example, a harpsichord fitted with a PHOTON system and connected via MIDI to a pipe organ can be configured as a hybrid harpsichord–organ hyperinstrument, in which a single or double historical manual simultaneously excites two physically distinct acoustic instruments. Rather than simply mirroring all played notes, the organ layer can be selectively engaged through performance-dependent mappings. For instance, a velocity-sensitive gate may be applied such that only fortissimo key inputs trigger the organ, while softer playing excites only the harpsichord. In this configuration, dynamic touch becomes a structural control parameter, allowing the performer to articulate fluidly between instrumental layers and to shape composite textures through embodied gesture rather than discrete switching actions.

\subsection{Ecologically Valid Reference for Digital-Instrument Research}
PHOTON enables systematic human–instrument interaction studies by providing high-resolution reference data from an actual historical keyboard mechanism. Digital pianos and MIDI controllers—whether used for research, practice, or performance—often suffer from the criticism that their keybeds feel “dead,” or unresponsive compared to acoustic instruments. By installing PHOTON inside a real harpsichord or fortepiano, it becomes possible to evaluate digital instrument software with an actual keybed. These measurements can serve as an upper bound or benchmark against which digital keybeds may be compared, informing the design of more lifelike digital actions and providing quantitative targets for future physical-modelling or haptic-feedback instruments.

Because PHOTON can be deployed inside an instrument without altering its behaviour, it also opens new methodological possibilities for studying performer biomechanics, motor learning, and expressive gesture in historically informed performance. Researchers can analyse how players adapt their technique across instruments, how articulation strategies correlate with key-motion profiles, or how tactile feedback influences interpretive choices—all within the ecological validity of a real, functioning instrument. Such studies have rarely been conducted on historical keyboards, and their biomechanics remain comparatively underdocumented.

\subsection{Pedagogical and Educational Applications}
Beyond performance and research contexts, PHOTON enables digitally augmented practice and rehearsal workflows on historical keyboard instruments—capabilities that have long been commonplace on digital pianos and MIDI controllers, but have not previously been available on harpsichords, clavichords, or other historical acoustic keyboards.
When combined with a virtual instrument, the PHOTON system allows the recording and replaying of music. This capability enables established practice techniques such as hands-separate practice, in which one hand is recorded and subsequently played back while the performer practices the other. It also supports four-hands or duet rehearsal on a single instrument, allowing one performer to practice against a recorded partner part. Because playback is rendered through virtual instruments, high-quality audio exports can be generated and shared, enabling the creation of accompaniment and rehearsal tracks for individual or ensemble practice. 

Finally, PHOTON's real-time MIDI output makes it possible to integrate historical keyboards directly with contemporary MIDI-enabled educational and practice software. 
Widely used platforms such as Simply Piano,\footnote{\url{https://www.joytunes.com/simply-piano}}, Yousician,\footnote{\url{https://yousician.com}} Piano Marvel,\footnote{\url{https://www.pianomarvel.com}} Skoove,\footnote{\url{https://www.skoove.com}} Flowkey,\footnote{\url{https://www.flowkey.com}} and power-user MIDI environments such as Midiano,\footnote{\url{https://midiano.com}} rely on precise note-on, note-off, and timing data to provide real-time feedback, assessment, and guided practice. 
Through such integration, students can practise repertoire, technique, and sight-reading on a historical instrument while benefiting from interactive pedagogical tools traditionally limited to digital pianos and MIDI controllers.

\subsection{Corpus Creation and Experimental Studies}
Beyond real-time applications, PHOTON enables the systematic creation of MIDI corpora captured directly from historical keyboard actions.
Unlike conventional MIDI datasets derived from digital pianos, generic MIDI controllers, or symbolic scores, recordings made with PHOTON encode performance data as it is shaped by the mechanical and organological context of specific historical instruments.

Such corpora make it possible to study articulation, timing, ornamentation, and expressive gesture as they emerge from real historical actions, rather than as idealised or cleaned symbolic representations.
This supports analyses that jointly consider symbolic structure and embodied interaction, enabling comparison across instruments with differing actions, key geometries, and sounding mechanisms.

Because captured performances can be replayed through configurable virtual instruments, the same underlying gestural data can be rendered under systematically varied synthesis conditions.
This enables controlled A--B experiments in which physical or acoustical parameters---such as temperament, tuning systems, or timbral models---are modified while performer input remains constant, allowing expert harpsichord performance to serve as a stable reference for studying the perceptual and musical consequences of these variables.

\subsection{Musically-Derived Performance Augmentation}
Beyond sound generation, the MIDI output of the PHOTON system can also serve as a real-time musical information stream for external systems such as adaptive stage or ambient lighting. By analysing performance parameters such as harmonic content, note density, articulation, and dynamic level, visual environments can be coupled directly to the performer’s gestures. In this way, keyboard performance can drive responsive lighting or stage systems, enabling visual augmentation that emerges from musical action rather than from preprogrammed cues.

\section{\texorpdfstring{Harpsichord Key–Action Behaviour:\\ A PHOTON Case Study}{Harpsichord Key–Action Behaviour: A PHOTON Case Study}}\label{sec:case_studies}
This case study examines how plucking events can be inferred from harpsichord key–action behaviour, providing a mechanically grounded perspective on touch that complements prior acoustic analyses.

The harpsichord is often described as having a limited dynamic range compared to the piano; however, experimental studies show that small but measurable dynamic and timbral differences are under the performer’s control, with reported level differences of up to about 11~dB across touch conditions even on a single manual and without changes to stops or registers \cite{penttinen2006dynamics,macritchie2015}. 
Although this range is modest relative to the piano, it indicates that meaningful dynamic variation exists and that the estimation of excitation velocity is not merely an abstract or musically irrelevant exercise. 
Prior studies have primarily characterized the acoustic consequences of different touches, and have also discussed historically situated schools of harpsichord technique in which distinct hand gestures and modes of key attack are associated with different tonal outcomes \cite{macritchie2015}. 
However, empirical investigations of the harpsichord remain comparatively sparse, particularly with respect to direct measurements of key motion. 

Unlike a typical digital piano, where velocity is often derived from the time delay between two staggered contact switches, measuring dynamics on a harpsichord presents distinct challenges.
A harpsichord key can rest in light contact with the string before plucking, during which the plectrum (and jack) encounters resistance from the string that is perceptible to the player.
The effective strike point does not occur near the end of the keystroke, but rather approximately midway through the key’s travel.
On double-manual instruments, the manuals exhibit slightly different strike positions due to intentional staggering.

\begin{figure}[t]
    \centering
    \includegraphics[width=1.0\linewidth]{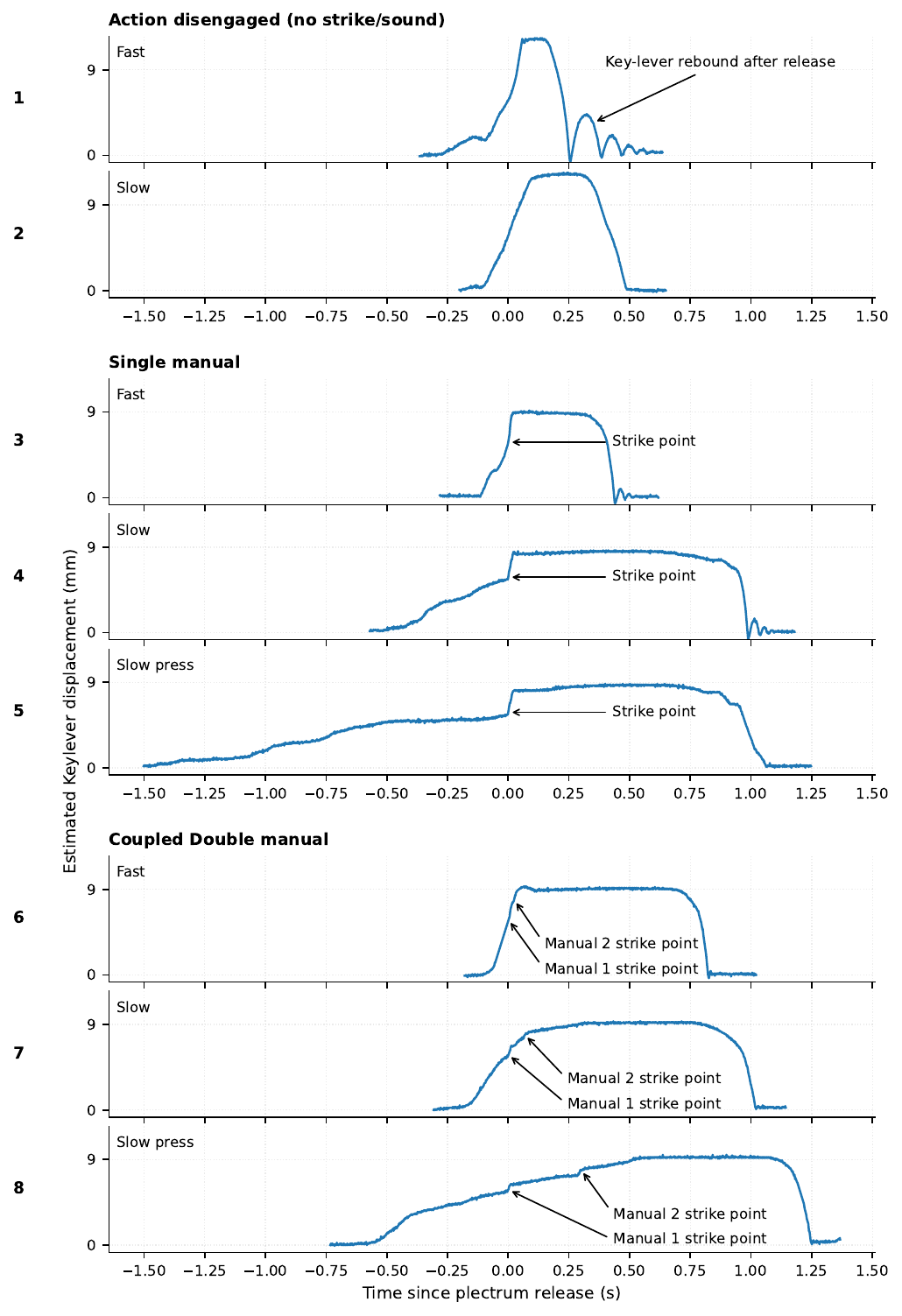}
\caption{Representative PHOTON motion traces showing key lever position over time under three configurations: disengaged single manual, single manual, and double manual. Traces are aligned to the pluck/strike event and share a common time and vertical scale.}
\Description{Eight stacked time-series plots of key position in Key Position versus time. Traces are grouped into three sections with labels; displacement crossing, numbered 1 through 8 on the left, and share consistent axes for comparison.}
    \label{fig:case_study_traces}
\end{figure}

Figure~\ref{fig:case_study_traces} presents PHOTON motion traces illustrating these interaction regimes across disengaged, single-manual, and double-manual configurations.
The motion traces were recorded on a double-manual Franco-Flemish harpsichord by G.C. Klop (1973) using PHOTON prototype hardware with a sampling rate of 250~Hz. 
In the single-manual traces, the effective pluck point is visible as a consistent feature near the position corresponding to approximately 5.5\,mm on the displacement axis, while in the double-manual configuration an additional, higher feature appears near 7.0\,mm, corresponding to the second plucking event produced by the staggered, coupled actions.
The displacement axis is obtained by mapping per-sensor ADC readings between two mechanically measured reference positions, with rest defined as 0\,mm displacement and full key travel measured as 9\,mm; intermediate values therefore form a monotonic but nonlinearly scaled coordinate rather than an exact geometric measurement.
Traces~1 and~2 show the key lever operating without engagement of any sound-producing mechanism, providing a reference for the unloaded mechanical response. 
In traces~1,~3, and~4, the key is released rapidly after depression, and the subsequent oscillatory settling of the key lever is visible as it returns to rest. 
Across the plucked conditions, the region of motion most relevant to excitation is localised around the pluck point(s), supporting the use of a narrower spatial window than is typical in digital-piano-style velocity estimation.
At the moment of plucking, the jack releases from the string and the key lever undergoes a brief unloading, producing a characteristic change in slope in the position trace. 
This feature reflects a release of mechanical resistance rather than the excitation force itself.
PHOTON derives both note-on and note-off velocity from measured event timing, but note-on is based on strike timing and note-off is based on release timing; in both cases the main host maps the measured time to MIDI velocity with the same configurable curve.
Note-off velocity is musically relevant on the harpsichord, since key release governs how quickly the jack returns and how the plectrum passes back across the string, a process that may produce audible release noise.

\section{Hardware Design}\label{sec:hardware} 
\subsection{Sensing Mechanism}
\begin{figure}
    \centering
    \includegraphics[width=1.0\linewidth]{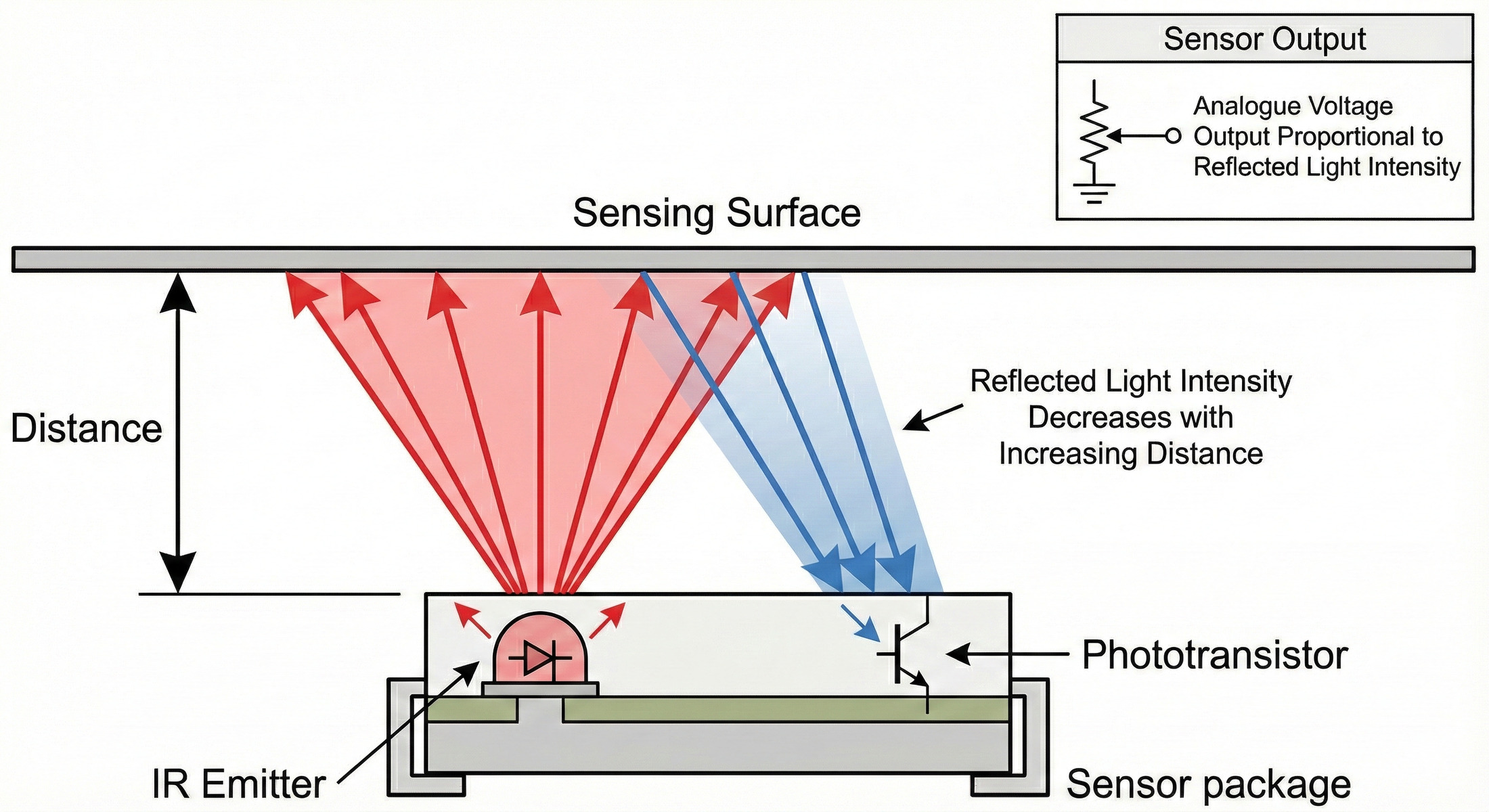}
    \caption{Sensing mode of the VCNT2025X01 sensor.}
    \Description{Diagram of a reflective optical sensor showing an infrared emitter and a phototransistor housed in a single package. Light is emitted toward a nearby surface, reflected back with intensity varying by distance and surface reflectivity, and received by the phototransistor.}
    \label{fig:reflective_sensor}
\end{figure}

The VCNT2025X01 is a compact reflective optical sensor manufactured by Vishay, integrating an infrared emitter and a phototransistor in a 2.5\,mm surface-mount package.\footnote{\url{https://www.vishay.com/en/product/84895/}} 
It belongs to the same class of reflective infrared sensors as devices such as the QRE1113 used in earlier continuous key-motion sensing systems, but in a smaller surface-mount form factor \cite{McPherson2013ContinuousPianoGesture,hamilton2025harpsichord}. 
When driven, the emitter projects infrared light onto a nearby surface; the phototransistor responds to the reflected radiation, producing an analogue voltage that depends on distance and surface reflectivity.
Optimised for short-range reflective sensing, the device is well suited for detecting key levers and other millimetre-scale displacements within a harpsichord mechanism.
Its low profile, high sensitivity, and minimal external component requirements allow it to be densely tiled across a sensor board while fitting comfortably inside the instrument.

Figure~\ref{fig:reflective_sensor} illustrates the operating principle.
Infrared light emitted by the device reflects off the underside of a key lever or jack and returns to the phototransistor.
The reflected intensity decreases smoothly with distance, producing a stable analogue signal suitable for tracking small vertical motions.
This reflective geometry avoids the alignment challenges of beam-break configurations, tolerates slight lateral motion, and permits the sensor to be mounted close to the moving component without altering the instrument.

\subsection{Schematics and PCB Design}
The VCNT2025X01 requires only a small number of supporting components.
A resistor on pin~3 sets the emitter brightness; an N-channel MOSFET between pin~4 and ground provides on--off control so that the sensor draws power only when enabled.
A pull-down resistor on the enable line ensures predictable off behaviour when the line is not driven.
A pull-up resistor on pin~2 (Y\_OUT) establishes the output voltage swing and sensitivity.
Y\_OUT is read by the microcontroller’s analogue-to-digital converter (ADC) to determine the presence and position of a reflective surface.

\begin{figure}[h]
    \centering
    \includegraphics[width=1.0\linewidth]{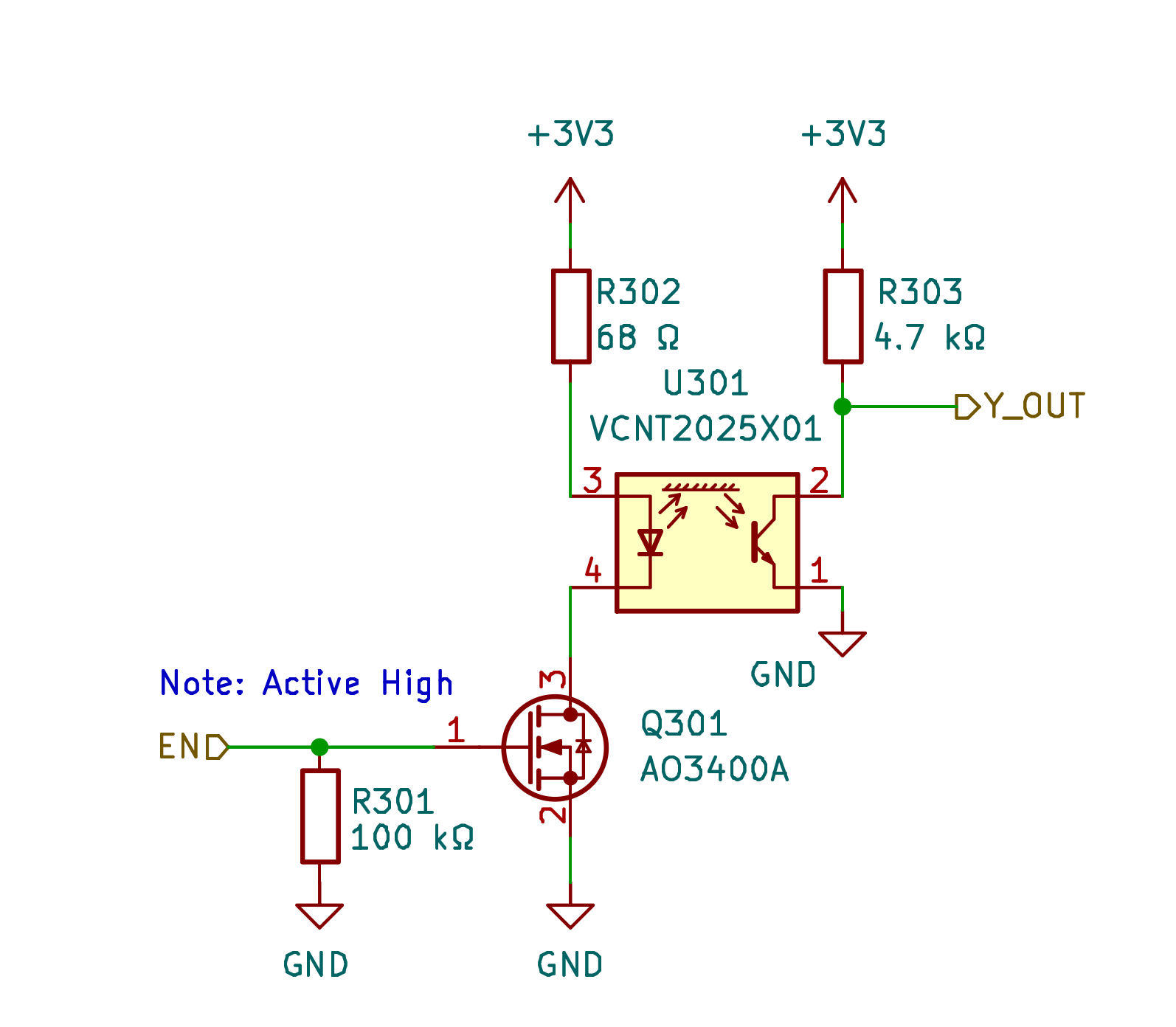}
    \caption{Schematic excerpt from the sensor array, showing a single sensor channel.}
    \Description{Schematic excerpt showing a single sensor channel built around a VCNT2025X01 reflective optical sensor, with supporting resistors and transistor. The diagram includes an enable control line and an analogue output signal.}
    \label{fig:digizer_array_schematic}
\end{figure}

\begin{figure}[h]
    \centering
    \includegraphics[width=1\linewidth]{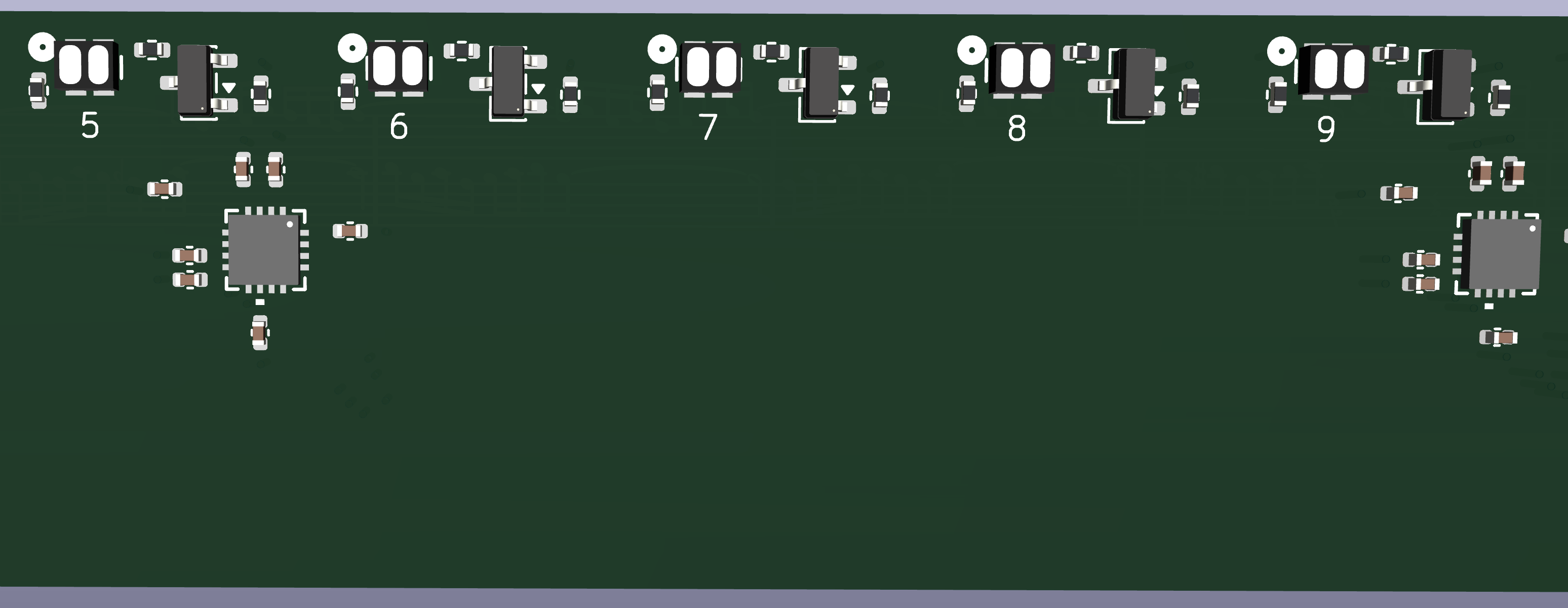}
    \caption{Printed circuit board assembly (PCBA) layout excerpt of the sensor array board, corresponding to Fig.~\ref{fig:digizer_array_schematic} and showing channels~5--9. Each channel comprises a VCNT2025X01 reflective optical sensor and minimal supporting components. The board width (Y dimension) is 25~mm.}
    \Description{Close-up photograph/render of the sensor array PCB assembly showing several repeated sensor channels and surface-mount components.}
    \label{fig:digitzer_array_layout}
\end{figure}

Each sensor belongs to a bank of four sensors associated with a Texas Instruments TLA2518 analogue-to-digital converter (ADC). 
Sensors are enabled one at a time using general-purpose input/output (GPIO) lines driven directly from the TLA2518 to the sensor enable pins. 
Measurements are taken by sampling the sensor Y\_OUT voltage using the ADC inputs of the TLA2518.
Communication with the TLA2518 uses a high-speed Serial Peripheral Interface (SPI), allowing high-throughput, low-latency sensor acquisition while retaining the accessibility and rapid-development advantages of CircuitPython on the microcontroller.

\subsection{System Integration and Design Rationale}
\label{sec:system_integration}
The architecture of the PHOTON sensing system reflects a sequence of design decisions made in direct response to the physical, mechanical, and electrical constraints described earlier.
The system is designed to operate across keyboards of arbitrary compass and manual configuration, and to function seamlessly as multiple coordinated modules.

Because common economical PCB fabrication and assembly services typically limit board dimensions to approximately 40--50\,cm in at least one direction,\footnote{Representative maximum assembled dimensions from several manufacturers at the time of writing include: PCBWay: 250×500\,mm, JLCPCB: 400×500\,mm, ALLPCB: 250×500\,mm, AISLER: 500×500\,mm, OSH Park: 406×558\,mm, and Eurocircuits: 340×440\,mm.} the system adopts a modular architecture in which identical sensor boards are placed side by side beneath the manual, allowing coverage of keyboards of arbitrary compass while maintaining a single, economical PCB design.
To accommodate manuals with odd numbers of keys, each board includes a removable end segment.
To avoid interference with the key action, the assembled board height is constrained to remain below 5.0\,mm, motivating the exclusive use of surface-mount components and low-profile connectors.

The PHOTON system is designed to be mechanically non-invasive and fully reversible.
In typical installations, sensor boards can be secured using a reversible, non-permanent mounting compound (e.g., removable adhesive putty such as Blu-tack), allowing placement beneath the key levers without drilling, screwing, or altering the instrument; for applications requiring greater mechanical robustness—such as long-term installations or transport—the boards also provide optional mounting holes for discreet fastening to existing structural elements of the keybed.

Reflective optical sensing is sensitive to surface reflectivity and mechanical variability between keys and instruments; to ensure consistent measurements across sensors and installations, PHOTON employs per-sensor calibration and normalisation.
During calibration, each sensor records reference values at rest and at full key travel, which are stored and used to normalise subsequent measurements. Event thresholds and velocity mappings are derived from these calibrated ranges, and calibration can be repeated whenever boards are repositioned or the instrument setup changes.

Historical keyboards exhibit substantial variability in key width, dip depth, and overall compass.
To align the sensing system with these differences, this project includes a custom KiCad plugin that automatically positions the optical sensors according to the measured distances between the guide pins located at the center of each key lever.
This allows the sensor array to match the exact key spacing of a given instrument while still relying on a shared underlying board architecture.
Modern small-batch PCB fabrication further supports this workflow by making bespoke boards practical and economically accessible.

The integration of an RP2350 microcontroller on each sensor board enables local acquisition and processing of high-rate sensor data, reducing latency and interconnect demands. Power and differential communication are provided via paired four-pin connectors, allowing boards to be daisy-chained in either direction.

A dedicated main board acts as the system host, coordinating communication and aggregating event data from the sensor boards. During operation, sensor boards perform local scanning and event detection and transmit event messages asynchronously.
The main board maintains system state and interfaces with external hosts and peripherals, and supports periodic status queries and explicit polling modes for diagnostics and calibration. Each sensor board stores a firmware configuration file encoding its identity and address, and includes a USB-C connector for direct firmware updates.

The system includes a main controller board that coordinates the distributed sensor modules and interfaces with external hosts and peripherals. 
The main board is powered via USB-C and distributes regulated power to the sensor modules, whose low idle and operating consumption permits extended battery-powered use without thermal or power-management constraints. In a 124-sensor, 5-board configuration, the complete system measured approximately 0.75~ A at 5~ V (3.75~ W) during operation, making battery-powered use from standard USB sources and portable power banks practical.

The system employs a differential RS-485 link rather than I\textsuperscript{2}C, SPI, or single-ended UART, as it provides a robust two-wire bus suitable for long cable runs and modular daisy-chaining. This supports reliable communication across multiple sensor boards while enabling high-speed data rates. All exposed pins incorporate electrostatic discharge (ESD) protection.

Taken together, these design decisions form a system that meets the mechanical constraints of historical harpsichords, the electrical and timing demands of high-resolution key-motion capture, and the practical limitations of modern PCB manufacturing.
The PHOTON architecture is optimised for real-world deployment rather than laboratory-only conditions.

\begin{figure}[t]
    \centering
    \includegraphics[width=1.0\linewidth]{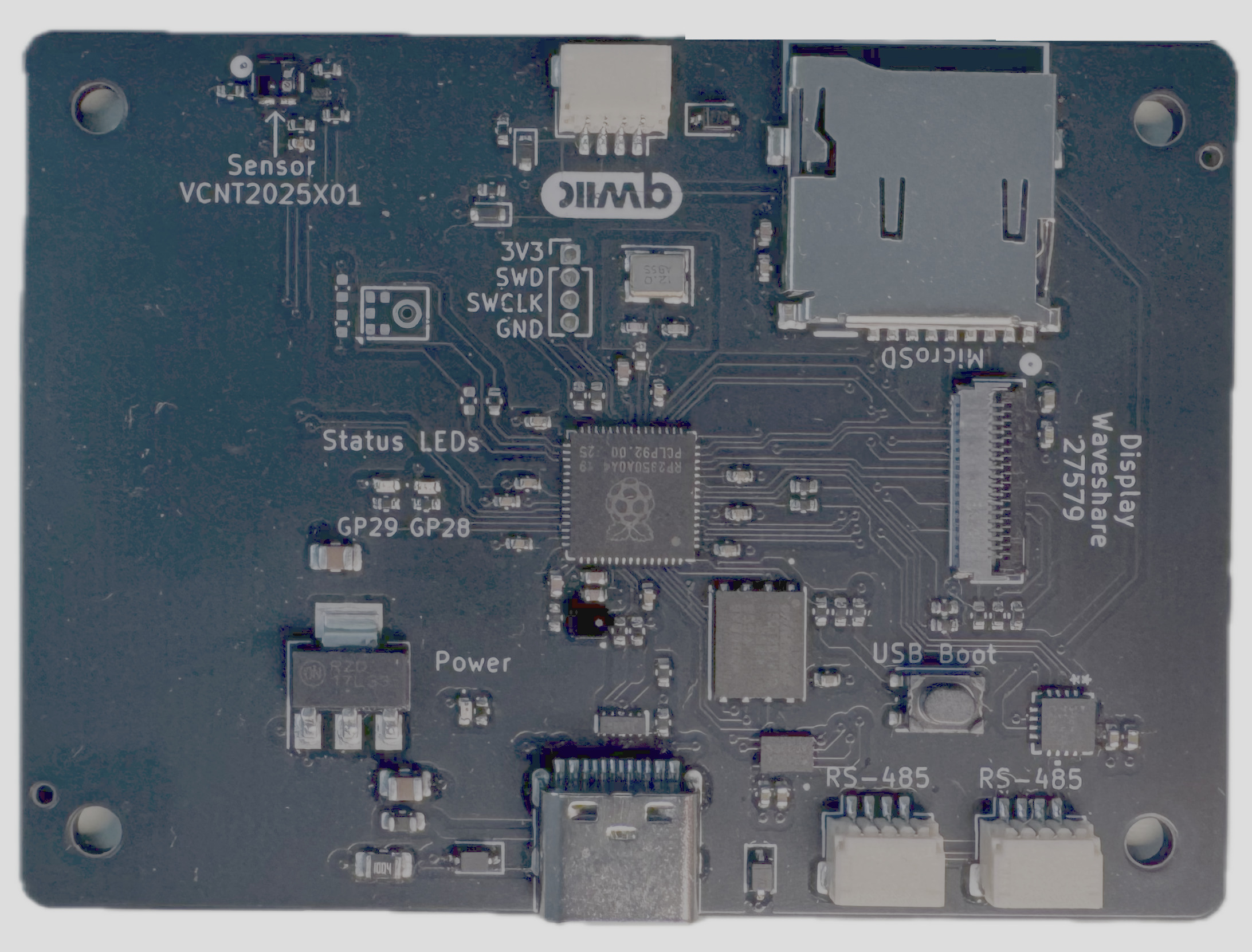}
    \caption{PHOTON main controller board featuring the RP2350 microcontroller. The board (approximately 7\,cm $\times$ 5\,cm) coordinates communication with the distributed sensor boards over an RS-485 bus and serves as the central processing and control unit of the system.}
    \Description{Photograph of the PHOTON main controller board showing the RP2350 microcontroller and supporting components, used to coordinate sensor modules and manage system communication.}
    \label{fig:mcu_board}
\end{figure}

\subsection{Extensibility}
To support experimentation and future expansion, the main board incorporates an I\textsuperscript{2}C bus with a standardised 4-pin JST-SH connector (compatible with Qwiic / STEMMA QT ecosystems), enabling rapid prototyping of additional sensors or auxiliary modules without redesigning the underlying hardware platform.\footnote{See \url{https://www.sparkfun.com/qwiic} and \url{https://learn.adafruit.com/introducing-adafruit-stemma-qt} for representative descriptions of these ecosystems.}
The board also includes external memory (PSRAM), a microSD card slot, and a touch-screen interface, which together enable fully standalone data acquisition and playback.
In particular, the system can record MIDI data directly to the microSD card, display real-time key-motion visualisations, and perform simple on-board analyses such as chord identification or articulation classification.
While not required for operation, a touch screen can also provide immediate feedback during practice or research sessions, for example by visualizing whether the performer is achieving desired techniques, such as resting the plectrum lightly on the string before plucking.

The system is powered via a USB-C interface and can operate from any standard USB power source, including portable battery packs, allowing fully self-contained, battery-powered use without the need for a host computer.
This capability, combined with the compact form factor, allows the entire system to remain concealed within the instrument when desired.
In such configurations, care must be taken to address the ethical implications of unobtrusive sensing; however, the design ensures that performers are not burdened with additional equipment and that their natural approach to the instrument remains unaltered.

\subsection{Reproducible Fabrication and Cost}
To support reproducibility and low-barrier fabrication, the full electronic design is provided with part numbers embedded directly in the schematics.
All components are annotated with LCSC part numbers, allowing the bill of materials to be resolved unambiguously and supporting automated assembly through standard small-batch commercial PCB manufacturing workflows.

For a typical small research deployment, fabrication of five identical sensor boards represents a practical and cost-effective batch size.
Table~\ref{tab:cost_estimate} summarizes a rough order-of-magnitude bill of materials for such a batch, including PCB fabrication and assembly including typical shipping, customs, and taxes.
At this scale, the per-board cost is dominated by three nearly equal components: setup costs, electronic components, and shipping and customs fees.

\begin{table}[t]
\caption{Indicative fabrication costs for PHOTON hardware (including EU delivery and import fees).}
\centering
\begin{tabular}{@{}l r@{}}
\toprule
\multicolumn{2}{@{}l}{\textbf{Sensor Boards (Batch of 5)}} \\
\midrule
Printed Circuit Boards            & \$16.70 \\
Components and Assembly           & \$190.56 \\
Cables                            & \$5.00 \\
Shipping, Customs, and Fees       & \$85.82 \\
\midrule
\textbf{Total}                    & \textbf{\$298.08} \\
Per-Board Cost                    & \$59.62 \\
\addlinespace[0.8em]

\multicolumn{2}{@{}l}{\textbf{Main Controller Boards (Batch of 5)}} \\
\midrule
Printed Circuit Boards            & \$4.20 \\
Components and Assembly           & \$120.00 \\
Misc.\ Hardware                   & \$5.00 \\
Shipping, Customs, and Fees       & \$72.44 \\
\midrule
\textbf{Total}                    & \textbf{\$201.64} \\
Per-Board Cost                    & \$39.24 \\

\addlinespace[1.0em]
\midrule
\addlinespace[0.4em]

\textbf{Grand Total}              & \textbf{\$499.72} \\
\bottomrule
\end{tabular}

\label{tab:cost_estimate}
\end{table}

To streamline fabrication, the project uses the open-source \emph{Fabrication Toolkit} plugin for KiCad, which can automatically generate Gerber files, pick-and-place (POS) data, and assembly-ready bills of materials directly from the design files.\footnote{\url{https://github.com/bennymeg/Fabrication-Toolkit}}
Together, the validated schematics, PCB layouts, and tested fabrication outputs enable reliable system reproduction without repeated design iteration.

\subsection{Open-Source and Accessible Technology}

A guiding principle of this project is the use of open-source tools and accessible hardware platforms to ensure long-term reproducibility and ease of adoption.
The electronic design workflow is based on KiCad, an open-source and free PCB design suite with transparent file formats and a robust scripting interface.

At the hardware level, the project relies on the RP-series microcontroller family, developed by Raspberry Pi.
The RP platform is positioned as a low-cost, widely available microcontroller ecosystem with excellent documentation, long-term supply commitments, and a thriving community.
By relying entirely on free, open-source software and widely available microcontroller hardware, the system remains easy to reproduce, modify, and extend.

Firmware development is conducted primarily in CircuitPython, an open-source, freely distributed framework designed to be far more accessible to newcomers than the lower-level Raspberry Pi Pico SDK.\footnote{The PHOTON hardware is also compatible with the Pico SDK}
CircuitPython allows rapid iteration, interactive debugging, and firmware updates without specialised toolchains.

To support the PHOTON platform, we provide a minimally modified fork of CircuitPython that includes board definition files for the PHOTON sensor and main controller boards, as well as a dedicated RS-485 driver tailored to the system’s event-driven communication model.
Precompiled firmware images for these boards are provided alongside the source code, allowing users to deploy the system without building CircuitPython.

\section{Online Resources}
All materials required to inspect, reproduce, and operate the PHOTON system are available at \url{https://github.com/w4iei/photon}.
These include:
\begin{itemize}
    \item electronic schematics and printed circuit board layouts (KiCad);
    \item custom KiCad plugin for sensor placement and board generation;
    \item CircuitPython firmware source code, including the RS-485 communication driver;
    \item precompiled firmware images for the sensor and main controller boards; and
    \item CircuitPython code used to operate the system
\end{itemize}

\section{Ethical Standards}
\balance
This paper presents a sensing system designed for the capture of performance data from keyboard instruments.
No formal human-participant study was conducted as part of this work.
All example measurements shown in the paper are derived from the author’s own interaction with the instrument.

Because the system can operate in a compact, self-contained, and concealed configuration, its use raises ethical considerations regarding unobtrusive sensing.
The system is intended for use in contexts where performers are informed of the system’s presence and purpose, and where appropriate consent has been obtained.

The authors report no conflicts of interest and no external funding that would influence the design, interpretation, or presentation of this work.

\bibliographystyle{ACM-Reference-Format}
\bibliography{digitizer}

\end{document}